\documentclass[prb,twocolumn,superscriptaddress,showpacs]{revtex4-1}
\usepackage{graphicx}
\usepackage{dcolumn}
\usepackage{bm}
\usepackage{color}
\usepackage{amsmath}
\usepackage{amssymb}
\usepackage{multirow}

\begin{document}

\title{Circularly Polarised X-ray Scattering Investigation of Spin-Lattice\\
Coupling in TbMnO$_3$ in Crossed Electric and Magnetic Fields}

\author{H. C. Walker}
\affiliation{ISIS Facility, Science and Technology Facilities Council, Rutherford Appleton Laboratory, Didcot, Oxfordshire OX11 0QX, UK}
\affiliation{European Synchrotron Radiation Facility, Bo\^{i}te Postale 220, 38043 Grenoble, France }
\affiliation{Deutsches Elektronen-Synchrotron DESY, D-22607 Hamburg, Germany}
\author{F. Fabrizi}
\affiliation{European Synchrotron Radiation Facility, Bo\^{i}te Postale 220, 38043 Grenoble, France }
\affiliation{London Centre for Nanotechnology, University College London, 17-19 Gordon Street, London WC1H 0AH, UK}
\affiliation{Diamond Light Source Ltd, Harwell Science and Innovation Campus, Didcot, Oxfordshire, OX11 0DE, UK}
\author{L. Paolasini}
\affiliation{European Synchrotron Radiation Facility, Bo\^{i}te Postale 220, 38043 Grenoble, France }
\author{F. de Bergevin}
\affiliation{European Synchrotron Radiation Facility, Bo\^{i}te Postale 220, 38043 Grenoble, France }
\author{D. Prabhakaran}
\affiliation{Department  of Physics, Clarendon Laboratory, University of Oxford, UK}
\author{A. T. Boothroyd}
\affiliation{Department  of Physics, Clarendon Laboratory, University of Oxford, UK}
\author{D. F. McMorrow}
\affiliation{London Centre for Nanotechnology, University College London, 17-19 Gordon Street, London WC1H 0AH, UK}

\date{\today}
\begin{abstract}
We present a study of the magnetic and crystallographic structure of TbMnO$_3$ in the presence of crossed electric and magnetic fields using circularly polarised X-ray non-resonant scattering. A comprehensive account is presented of the scattering theory and data analysis methods used in our earlier studies, and in addition we present new high magnetic field data  and its analysis.
We discuss in detail how polarisation analysis was used to reveal structural information, including the arrangement of Tb moments which we proposed for $H = 0$~T, and how the diffraction data for $H<H_C$ can be used to determine specific magnetostrictively induced atomic displacements with femto-metre accuracy. The connection between the electric polarisation and magnetostrictive mechanisms is discussed. Similar magnetostrictive displacements have been observed for $H > H_C$ as for $H < H_C$.
Finally some observations regarding the kinetics and the conservation of domain population at the transition are described.
\end{abstract}

\pacs{75.25.-j, 75.47.Lx, 75.85.+t, 77.80.Dj}

\maketitle

%
%

\section{INTRODUCTION}

The magnetoelectric effect, as postulated in 1894 by Pierre Curie\cite{Curie}, in which magnetic and electric order are coupled together, is obviously a promising candidate for all manner of technological applications\cite{Fiebig}. However, for some forty years after the first experimental observation\cite{Astrov}, it proved to be little more than a diverting scientific curiosity thanks to only a very weak coupling, as a consequence of the limited magnetic and electric susceptibilities. Therefore, the discovery of a giant magnetocapacitance effect in TbMnO$_3$\cite{KimuraTMO} in 2003, and the associated modern advent in multiferroics, has led to a seismic shift in the theoretical and experimental research output in this field, as we have sought to obtain a fundamental understanding of the underlying physics. This is essential if we hope to fully exploit the potential for devices, such as magnetic computer memory which could be written using an electric field, which would be more energy efficient\cite{Bibes}.

TbMnO$_3$ is not a proper ferroelectric, in which a structural instability towards the polar state, associated with the electronic pairing, would be the main driving force behind the spontaneous polarisation. Instead it is an improper ferroelectric, since the polarisation appears as an accidental by-product of some other ordering, in this case magnetic ordering, as implied by the onset of a finite spontaneous ferroelectric polarisation concomitant with a magnetic phase transition at $T=27$~K\cite{KimuraTMO}. Further insight into the driving mechanism was obtained when neutron diffraction\cite{kenzelmann} determined that the incommensurate magnetic structure goes from a sinusoidal ordering (non-polar) to one described by a cycloid formed by the Mn$^{3+}$ ion spins, which breaks inversion symmetry and makes TbMnO$_3$ polar, at the multiferroic phase transition.

A similar magnetically driven multiferroic mechanism is present in Ni$_3$V$_2$O$_8$\cite{Lawes, nivo} and MnWO$_4$\cite{Taniguchi}, and these materials are collectively classified as cycloidal Type II multiferroics\cite{Khomskii}. The resulting strong magnetoelectric coupling in TbMnO$_3$ is demonstrated by the possibility of controlling the magnetic domain state using an electric field \cite{Fabrizi, *FabriziE, yamasaki, *yamasakiE}, and by the switching of the ferroelectric polarisation axis from $\hat{\mathbf{c}}$ to $\hat{\mathbf{a}}$ on application of a magnetic field along the $\hat{\mathbf{b}}$ axis\cite{KimuraTMO}. Given the clear significance of the magnetic structure of TbMnO$_3$ to its magnetoelectric properties, it has been studied in detail using neutrons\cite{kenzelmann,yamasaki,*yamasakiE,AliouanePRL}, resonant X-ray scattering\cite{mannix,Strempfer,Forrest,Wilkins,LoveseyTMO}, and non-resonant X-ray scattering\cite{Fabrizi,*FabriziE,WalkerScience}.

However, despite all of this progress, until recently\cite{WalkerScience} the details of the microscopic mechanism driving the ferroelectricity remained obscured. Two different schools of thought had developed: one invoked a purely electronic mechanism, where the spin-orbit interaction modifies the hybridisation of the electronic orbitals to generate an electric polarisation\cite{Katsura2005}; whilst the other involves the lattice, where the Dzyaloshinskii-Moriya interaction leads to ionic displacements, and hence an electric polarisation\cite{Sergienko}. Ab-initio density function theory calculations supported the significance of the lattice\cite{Malashevich,Xiang}, but any displacements were beyond the experimental limit of EXAFS\cite{Bridges}; and it is only through recent developments in non-resonant X-ray scattering that these displacements have finally been measured\cite{WalkerScience}, providing conclusive support for the picture of Sergienko and Dagotto\cite{Sergienko}.

In this paper we provide a comprehensive overview of the physics behind the magnetoelectric coupling mechanism in TbMnO$_3$, as obtained through circularly polarised X-ray non-resonant scattering in crossed electric and magnetic fields. In so doing we present details that were omitted from our earlier papers\cite{Fabrizi,*FabriziE,WalkerScience} due to space limitations, as well as providing new data and analysis to complete our survey of the magnetic field--temperature phase diagram. The paper is set out as follows: Section~\ref{s:expdetails} explains the experimental technique, giving details of the X-ray polarisation control, and introduces the non-resonant magnetic scattering amplitude, giving a generic example for a cycloidal spin system, Section~\ref{s:H0} reviews the results in zero magnetic field, demonstrating how the technique is complementary to that of neutron diffraction for performing complex magnetic structure refinement, Section~\ref{s:Hn0} reviews the results in low applied magnetic field, and discusses how they may be taken in conjunction with a symmetry analysis to determine femtoscale ionic displacements, and finally Section~\ref{s:H10} presents new data in the high magnetic field polarisation flop phase.

%
%

\section{Non-resonant X-ray scattering}
\subsection{Theory}

The interaction between matter and the electric and magnetic fields of X-ray radiation results in a differential X-ray scattering cross-section composed of three terms:

\begin{equation}
\frac{d\sigma}{d\Omega}=r_0^2  \left |F_0+F_{NR}+F_{AN}\right|^2;
\end{equation}
where $F_0=\sum_n e^{i \mathbf{Q} \cdot \mathbf{r}_n} f_0$ is the charge structure factor which contains the Thomson x-rays scattering amplitude $f_0$, $F_{NR}$ is the non-resonant (NR) magnetic structure factor contribution and $F_{AN}$ is the anomalous (AN) or resonant contribution. The radiation interaction with the magnetic moments is much weaker, such that the intensity arising from pure magnetic diffraction is typically more than a factor of $10^6$ smaller than the intensity of classical Thomson charge scattering. It is for this reason, combined with the advantageous element specificity, that most X-ray investigations of magnetic order have instead used resonant scattering, exploiting the intensity enhancement observed at the absorption edges, which can be from a factor of $100$ up to $10^7$ at the rare-earth $L$-edges\cite{Gibbs} and actinide $M$-edges\cite{Isaacs}. However, at the $K$-edge of the $3d$ transition metals the resonant enhancement is typically only of order four \cite{Paolasini08}, such that at the absorption edge the scattered intensity will be composed of both resonant and non-resonant scattering processes, considerably reducing the benefits of performing X-ray resonant scattering experiments in this regime, because the non-resonant and resonant x-ray scattering amplitude interfere, being of the same order of magnitude. As a result, the interpretation of non-resonant magnetic scattering is more straight-forward than that of resonant magnetic scattering, since it requires no more assumptions than the spin and orbital parts of the ordered moment.

In this paper we exploit the unique direct coupling of non-resonant magnetic X-ray scattering to the magnetic structure,  which separates out the spin and orbital parts of the ordered moment, as revealed in the scattering amplitude for an isolated system of moments \cite{Blume, Altarelli}:

\begin{equation}
\label{eq:NRisolated}
F_{NR}=-i\frac{\hbar\omega}{mc^2}\left[\mathbf{L}(\mathbf{K})\cdot
\mathbf{P}_L + \mathbf{S}(\mathbf{K})\cdot\mathbf{P}_S\right],
\end{equation}

where $\mathbf{L}(\mathbf{K})$ and $\mathbf{S}(\mathbf{K})$ are the Fourier transforms of the atomic orbital magnetisation density and the spin density, and $\mathbf{P}_L$ and $\mathbf{P}_S$ are the polarisation factors defined as:

\begin{eqnarray}
\mathbf{P}_L&=&-\mathbf{\hat K}\times[\left(\boldsymbol{\hat\epsilon}^{\prime*}
\times\boldsymbol{\hat\epsilon}\right)\times\mathbf{\hat K}]\,2\sin^2\theta,
\label{P_L}\\
\mathbf{P}_S&=&%
\boldsymbol{\hat\epsilon}^{\prime*}\times\boldsymbol{\hat\epsilon}+\left(\mathbf{\hat k}^\prime\times
\boldsymbol{\hat\epsilon}^{\prime*}\right)\left(\mathbf{\hat k}^\prime\cdot
\boldsymbol{\hat\epsilon}\right)\nonumber \\
&-&\left(\mathbf{\hat k}\times
\boldsymbol{\hat\epsilon}\right)\left(\mathbf{\hat k}\cdot
\boldsymbol{\hat\epsilon}^{\prime*}\right)-\left(\mathbf{\hat k}^\prime\times
\boldsymbol{\hat\epsilon}^{\prime*}\right)\times\left(\mathbf{\hat k}\times
\boldsymbol{\hat\epsilon}\right)
\label{P_S}
\end{eqnarray}

\noindent where $\mathbf{k}$ ($\mathbf{k}^\prime$) is the incident (scattered) wave-vector and $\boldsymbol{\hat\epsilon}$ ($\boldsymbol{\hat\epsilon}^{\prime}$) is the incident (scattered) complex unit vector (in the case of circularly polarised x-rays), describing the x-rays polarisation state. Simple scattering theory implies that the resonant contribution decays slowly, only as the inverse of the difference between the resonance and the photon energy, such that even $5$~keV away from a large resonance one might expect a significant resonant scattering contribution.
Care should also be taken to take any quadrupolar ($E2-E2$) resonant scattering into account\cite{ScagnoliPRB,mannix}.
However, when a full analysis of the theory of magnetic scattering is performed \cite{Blume_malente} it is revealed that the resonant scattering length falls off faster than predicted. For example, for our experiments on TbMnO$_3$ performed at $6.16$~keV, the resonant contribution from the Tb $M_{IV}$ (E= 1276.9 keV) and $M_V$ (E=1241.1 keV) edges is calculated to be only $2-5\%$ of the non-resonant contribution, as  explained in the supporting online material to Ref.[\onlinecite{WalkerScience}]. Therefore, by moving away from the edge, it is possible to neglect the resonant term.

\subsection{Experimental Implementation}
\label{s:expdetails}

The experiments were performed at the former ID20 magnetic scattering beamline  of the European Synchrotron Radiation Facility in Grenoble \cite{Paolasini07}.
A single crystal of TbMnO$_3$, with dimensions $1\times1\times0.6$~mm$^3$ (Space group $Pbnm$, lattice parameters at room temperature $a_0$=5.3019 \AA, $b_0$=5.8557 \AA, $c_0$=7.4009 \AA) along the main crystallographic axes, was synthesized in Oxford using the floating zone method. The sample was glued between two copper plates using highly conductive silver paste to enable the application of an electric field of up to $2$~kV/mm along the $\hat{\mathbf{c}}$ direction. This assemblage was inserted into either an orange cryostat or the 10T Oxford Instruments cryomagnet giving an $\mathbf{a}-\mathbf{c}$ horizontal scattering plane with the potential to apply a magnetic field along the $\hat{\mathbf{b}}$ direction.
The experiments in zero magnetic field were performed at an incident X-ray energy of $7.45$~keV with an Au(222) polarisation analyser crystal. An orange cryostat was used in preference to the cryomagnet, so as to have access to a greater region of reciprocal space, including the complete star of wave-vectors (4,$\pm\tau$,$\pm1$).
The incident X-rays were converted from a linear to a circular polarisation state (typically $99\%$ circularly polarised) by a phase plate in the quarter-wave mode. A high quality ($1\bar{1}0$) diamond single crystal ($720$~microns thick) was used for this purpose, where the scattering was in Laue geometry from the ($111$) plane\cite{Scagnoli}. The sense of rotation of the circular polarisation state was defined according to the rule that for $\theta>\theta_B$ the polarisation rotates in the same direction as the $45^\circ$ rotation which brings the incident polarisation plane onto the scattering plane.

A complete Poincar\'e-Stokes polarimetry analysis\cite{deBergevin,Blume,Mazzoli} of the scattered beam ($\hat{\mathbf{k}^\prime}$) is obtained by collecting the dependence of the intensity $I(\eta)\propto1+P_1\cos(2\eta)+P_2\sin(2\eta)$ as a function of $\eta$, the rotation angle of a high quality analyser crystal about $\hat{\mathbf{k}^\prime}$. The Stokes parameter ${P}_1=[I(0^{\circ})-I(90^{\circ})]/ [I(0^{\circ})+I(90^{\circ})]$  is the polarisation rate parallel and perpendicular to the scattering plane defined by the analyser, and ${P}_2=[I(45^{\circ})-I(-45^{\circ})]/ [I(45^{\circ})+I(-45^{\circ})]$ the degree of oblique polarisation\cite{comment}. They can be calculated directly from the expression of the complex polarisation vectors $\hat\epsilon$:

 \begin{eqnarray}\label{equ:stokes}
  P_1 = \frac{ (\hat\epsilon_{\sigma} \hat\epsilon_{\sigma}^*-\hat\epsilon_{\pi} \hat\epsilon_{\pi}^*)} {(\hat\epsilon_{\sigma} \hat\epsilon_{\sigma}^*+\hat\epsilon_{\pi} \hat\epsilon_{\pi}^*)} , \  \  \
   P_2 = \frac{(\hat\epsilon_{\sigma} \hat\epsilon_{\pi}^*+ \hat\epsilon_{\pi} \hat\epsilon_{\sigma}^*) }{(\hat\epsilon_{\sigma} \hat\epsilon_{\sigma}^*+\hat\epsilon_{\pi} \hat\epsilon_{\pi}^*)}.
\label{eq:Poincare-Stokes1}
\end{eqnarray}

\begin{figure}[!htb]
\includegraphics[width=.9\linewidth, angle=0]{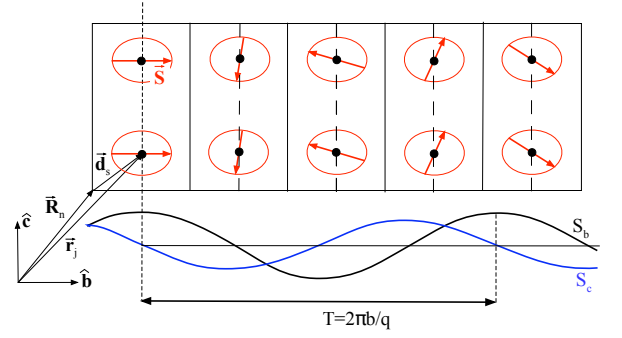}
\caption{A generic incommensurate spin cycloid (red arrows)  in the $\mathbf{b}-\mathbf{c}$-plane, propagating along $\hat{\mathbf{b}}$ directions with periodicity $T=2\pi b/q$, could be described by a magnetic vector $\vec{\mathbf{S}}$, decomposed in two sinusoidal orthogonal components $S_b$ and $S_c$ in quadrature, along the crystallographic axis $\mathbf{\hat b}$ and  $\mathbf{\hat c}$, respectively.}
\label{Fig:cycl}
\end{figure}

\subsection{Calculation of the structure factor for a spin cycloid}
For example, let we consider a generic incommensurate cycloidal spin structure in an orthorhombic cell (see Fig.~\ref{Fig:cycl}), in which there is a phase shift of $\pi/2$ between two orthogonal components: $S_b\hat{\mathbf{b}}$ and $S_c\hat{\mathbf{c}}$ which propagate according to $\mathbf{q}=2\pi \tau\mathbf{b}^\ast=\frac{2\pi\hat{\mathbf{b}}}{T}$, where $T$ is the magnetic period. Notice that $\mathbf{S}$ contains the $\mathbf{K}$ dependence of the magnetic spin form factor, and we suppose {\bf L}=0.

Then the incommensurate modulated spin moment may be written as:

\begin{equation}\label{eq:spins}
\begin{split}
\mathbf{S}_j&=S_b\cos(\mathbf{q\cdot r}_j)\hat{\mathbf{b}}+S_c\cos(\mathbf{q\cdot r}_j+\pi/2)\hat{\mathbf{c}}\\
&=\frac{1}{2}e^{i\mathbf{q\cdot r}_j}\left[S_b\hat{\mathbf{b}}+iS_c\hat{\mathbf{c}}\right] + \frac{1}{2}e^{-i\mathbf{q\cdot r}_j}\left[S_b\hat{\mathbf{b}}-iS_c\hat{\mathbf{c}}\right]\\
&=\frac{1}{2}e^{i\mathbf{q\cdot r}_j}\mathbf{M}+\frac{1}{2}e^{-i\mathbf{q\cdot r}_j}\mathbf{M}^\ast
\end{split}
\end{equation}

\noindent where $\mathbf{M}=(S_b\hat{\mathbf{b}}+iS_c\hat{\mathbf{c}})$.
Inserting this into equation~\eqref{eq:NRisolated}, and neglecting the orbital contribution, we obtain:

\begin{equation}\label{eq:cycspins}
\begin{split}
F_{NR}(\mathbf{K}) & =
-\frac{i\hbar\omega}{2mc^2} \left\{ \sum_n e^{i\mathbf{(K+q)\cdot R}_n}\sum_s e^{i\mathbf{(K+q)\cdot d}_s}\mathbf{M} \right. \\
 & + \left. \sum_n e^{i\mathbf{(K-q)\cdot R}_n} \sum_s e^{i\mathbf{(K-q)\cdot d}_s} \mathbf{M}^\ast \right\} \cdot\mathbf{P}_S,
\end{split}
\end{equation}
where the position of the $j$th atom $\mathbf{r}_j$ has been decomposed into $\mathbf{R}_n$, defining the position of the $n$th cell to which the $j$th atom belongs, and $\mathbf{d}_s$, the position of the atom within the cell ($\mathbf{d}_s=x_s\mathbf{a}+y_s\mathbf{b}+z_s\mathbf{c}$).

By spanning the scattering vector $\mathbf{K}=\mathbf{k}-\mathbf{k}^\prime$\cite{notedef} on the basis defined by the reciprocal lattice it is found that diffraction occurs for discrete values of $\mathbf{K}\pm\mathbf{q}=h\mathbf{a}^\ast+k\mathbf{b}^\ast+l\mathbf{c}^\ast$, where $h$, $k$, $l$ are integer numbers (Laue's diffraction condition), giving rise to twin reflections:

\begin{equation*}
\begin{split}
\sum_n e^{i\mathbf{(K+q)\cdot R}_n}\neq0\rightarrow& \mathbf{K}=h\mathbf{a}^\ast+(k-\tau)\mathbf{b}^\ast+l\mathbf{c}^\ast\\
\sum_n e^{i\mathbf{(K-q)\cdot R}_n}\neq0\rightarrow& \mathbf{K}=h\mathbf{a}^\ast+(k+\tau)\mathbf{b}^\ast+l\mathbf{c}^\ast,
\end{split}
\end{equation*}

\noindent with the magnetic structure factors:

\begin{equation}
F_{NR}(h,k,l)\propto
\begin{cases}
\sum_s e^{2\pi i(hx_s+ky_s+lz_s)}\left(\mathbf{M\cdot P}_S\right)\\
\sum_s e^{2\pi i(hx_s+ky_s+lz_s)}\left(\mathbf{M^\ast\cdot P}_S\right)\\
\end{cases}
\end{equation}

Clearly the diffraction amplitudes differ for the two reflections, but whether the measured intensities will differ depends on the polarisation state of the incident beam via $\mathbf{P}_S$. For linear polarised X-rays $\mathbf{P}_S$ is real and so $(\mathbf{M}^\ast\cdot  \mathbf{P}_S)=(\mathbf{M}\cdot  \mathbf{P}_S)^\ast$, with the result that the intensities of the two reflections will be identical.
However, for circularly polarised X-rays $\mathbf{P}_S$ contains complex polarisation vectors, with the result that $(\mathbf{M}^\ast\cdot \mathbf{P}_S)\neq( \mathbf{M}\cdot \mathbf{P}_S)^\ast$, and so $F_{NR}(h,k-\tau,l))\neq F_{NR}(h,k+\tau,l)$. Moreover, the left circular polarisation (LCP) and the right circular polarisation (RCP) behaviour change symmetrically if the sense of rotation of the cycloid change ($\mathbf{M}\rightarrow \mathbf{M}^*$) or, equivalently, by considering the opposite magnetic satellite $-\mathbf{\tau}$. This reveals the power of circularly polarised X-rays for studying helical, chiral or cycloidal magnetic structures, and  allows the determination of magnetic domain population factor associated to the ratio between of clockwise or anti-clockwise cycloids \cite{nivo}.

\begin{figure}[!htb]
	\begin{center}
		\includegraphics[width=.85\linewidth, angle=0]{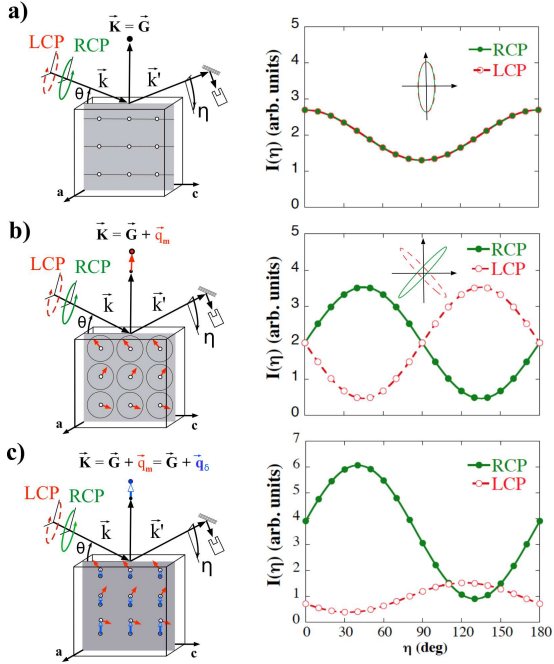}
	\end{center}
	\caption{
	a)  Circularly polarised x-ray diffraction of a Bragg reflection (Thomson scattering). The scattered x-ray intensities have the same $\eta$ dependence for RCP (green dots) and LCP (red open circles).  b) Circularly polarised x-ray magnetic diffraction of a satellite reflection associated with a magnetic cycloid propagating in the direction of the scattering vector ${\bf K}={\bf G}+{\bf q}_m$. c) Interference between an incommensurate displacement wave and the magnetic cycloid at the same ${\bf K}$ (${\bf q}_{m}={\bf q}_{\delta}$). The ratio between the intensities is $t\approx0.4$. (Adapted from Ref. \onlinecite{WalkerScience})
	}
	\label{fig:circular}
\end{figure}

Let us consider the simple case described in  Fig.\ref{fig:circular}, in which we compare the complete Poincar\'e-Stokes polarisation analysis of Thomson and magnetic diffracted intensities $I(\eta)$. The incident circular polarisation is described by the complex vectors $\hat{\boldsymbol\epsilon}_{RCP}= \left(\hat\epsilon_{\sigma} - i\hat\epsilon_{\pi}\right)$ and  $\hat{\boldsymbol\epsilon}_{LCP}= \left(\hat\epsilon_{\sigma} + i\hat\epsilon_{\pi}\right)$.
In order to calculate the Poincar\'e-Stokes parameters given in Eq.\ref{eq:Poincare-Stokes1}, we evaluate the final polarisation state ${\boldsymbol \epsilon}'$, exploiting the Jones matrix for the Thomson scattering (see Ref.\onlinecite{Paolasini08}):
\begin{equation}
    {\boldsymbol \epsilon}'
        = \left[
    \begin{array}{cc}
        1 & 0 \\
        0 & \cos 2\theta
    \end{array}
    \right]
\left[
    \begin{array}{c}
        1  \\
        \pm i
    \end{array}
    \right]
         =  \left[
    \begin{array}{c}
        1  \\
       \pm i \cos2\theta
    \end{array}
    \right]
    \label{eq:polthomsonscattering}
\end{equation}

\noindent and from Eq. \ref{eq:Poincare-Stokes1} we can calculate the Poincar\'e-Stokes terms ${P}_i$ for the scattered polarisations:

 \begin{eqnarray}
   P_1 = \frac{1- \cos^22\theta}{1+ \cos^22\theta}; \ \ \
   P_2 =  0.
   \label{eq:Poincare-Stokesthomson}
\end{eqnarray}

Both the circular polarisations give the same scattered polarisation dependence $I(\eta)$, as presented in Fig.\ref{fig:circular}(a), and when the Bragg angle $\theta=45^{\circ}$ the scattered light is completely vertically linear polarised, because $P_1\!=\!1$ and $P_2=0$. For a different scattering angle, the scattered polarisation is elliptical, because a circular contribution $P_3\!\neq\!0$ is present, but still vertical because $P_2\!=\!0$.

In the case of magnetic cycloid defined in Eq.\ref{eq:spins}, and supposing that the propagation vector ${\bf q}_m=2\pi \tau {\bf b}^*$ parallel to the scattering vector ${\bf K}$, we can calculate the Jones matrix from the magnetic scattering amplitude:

\begin{eqnarray}
    {\boldsymbol \epsilon}'  &=& 2S \sin^2\theta \left[
    \begin{array}{cc}
        0 & - (\cos\theta - i \sin\theta) \\
         (\cos\theta\!+ i \!\sin\theta) & 0
    \end{array}
    \right]
\left[
    \begin{array}{c}
        1  \\
        \pm i
    \end{array}
    \right]
      \nonumber    \\
  &  = & 2 S \sin^2\theta \left[
    \begin{array}{c}
        \mp \left( i \cos\theta +  \sin\theta\right)  \\
         \cos\theta + i  \sin\theta
    \end{array} \right]
    \label{eq:polmagncycloid}
\end{eqnarray}

\noindent where we have supposed that $S_b$=$S_c$=$S$. The Poincar\'e-Stokes can be calculated easily from Eq.\ref{equ:stokes}  :

 \begin{eqnarray}
{P}_1 =  0; \ \ \
{P}_2 =  \pm \sin 2 \theta.
   \label{eq:Poincare-Stokescycloid}
\end{eqnarray}

Now the scattered light has a dominant oblique polarisation $P_2\!=\!\pm \sin 2 \theta$, with the opposite sign for LCP and RCP, as shown in Fig.\ref{fig:circular}(b). When the scattering angle is close to $\theta\!=\!45^{\circ}$, it becomes completely linear and oblique ($P_2\!=\!1$ and $P_3\!=\!0$).

Finally, it is interesting to analyze the case in which there is an interference between the scattering arising from the incommensurate cycloidal magnetic structure and a displacement wave at the same propagation vector ${\bf q}_{\delta}={\bf q}_m$, for example as a consequence of the application of an external magnetic field (see Sec.\ref{s:Hn0}).
In this case the scattered intensities for RCP and LCP interfere and the Poincar\'e-Stokes parameters ${P}_i$ may be calculated by considering the complex sum of the  magnetic and Thomson Jones matrices.
After some easy calculation, the Poincar\'e-Stokes parameters may be determined:

\begin{eqnarray}
     P_1 &  = & \frac{t^2 (1-\cos^22\theta) \pm 2 t (1-\cos2\theta)\cos\theta}{ t^2 (1+\cos^22\theta) \pm 2 t  (1+\cos2\theta)\cos\theta+2}
    \nonumber \\
     P_2 & =  &  \frac{2t (1+\cos2\theta)\sin\theta \pm 2  \sin2\theta}{t^2 (1+\cos^22\theta) \pm 2 t  (1+\cos2\theta)\cos\theta+2}
    \label{eq:stokesinterf}
\end{eqnarray}

\noindent where $t$ is the ratio between the Thomson and magnetic scattering amplitude.
Fig.\ref{fig:circular}(c) shows the variation of $I(\eta)$ for LCP and RCP polarisations for a weak Thomson scattering contribution ($t=0.4$) superposed to the magnetic scattering, revealing the strong sensitivity of this method to the determination of small displacements associated to the magnetoelastic coupling, as we will demonstrate in Sec.\ref{s:Hn0}.
Notice that when $t\!\rightarrow\!\infty$ we obtain the case of Fig.\ref{fig:circular}(a) valid for the Thomson scattering, whereas when $t\!\rightarrow\!0$ the case of Fig.\ref{fig:circular}(b) for the pure magnetic scattering.

%
%

\section{Magnetic structure  at H=0 T in the incommensurate ferroelectric phase.}
\label{s:H0}

\begin{figure}[!htb]
\includegraphics[width=.9\linewidth, angle=0]{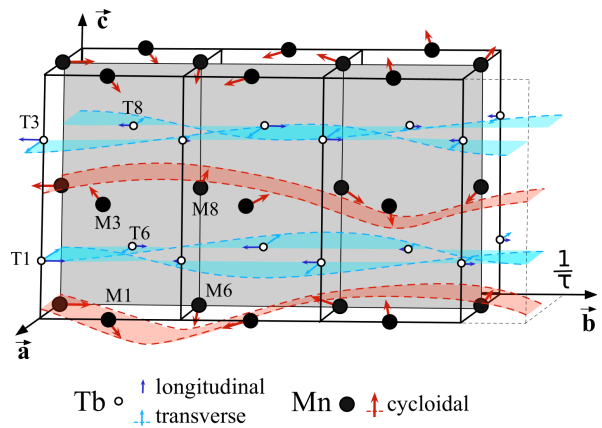}
\caption{Magnetic structure of TbMnO$_3$ determined at $T=15$~K. Only Tb and Mn atoms are shown in the crystallographic $Pbnm$ unit cell. The red (blue) arrows refers to the Mn (Tb) magnetic moment. (Modified from Ref. \onlinecite{Fabrizi,*FabriziE}).}
\label{Fig:structure}
\end{figure}

\begin{figure}[!htb]
\centering
\includegraphics[width=.45\textwidth]{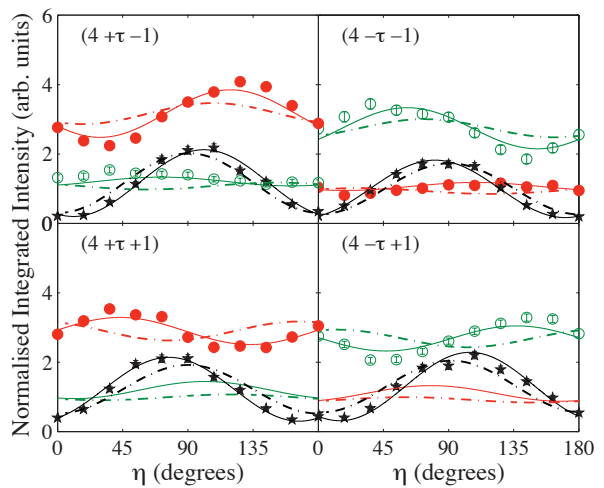}
\caption{\label{fig:H0stokes}Stokes dependence of the x-ray magnetic scattering in TbMnO$_3$ at $T=15$~K from the (4 $\pm\tau$ $\pm1$) reflections, after 
cooling in an electric field along $+\mathbf c$, measured with LCP (\textcolor[rgb]{1,0,0}{$\bullet$}), RCP (\textcolor[rgb]{0,0.59,0}{o}) and $\pi$ ($\star$) incident on the sample. The solid lines correspond to our model 
calculations, whilst the broken lines correspond to the model derived by Kenzelmann et al.\cite{kenzelmann} from neutron diffraction results (Adapted from Ref.\onlinecite{Fabrizi,*FabriziE}).}
\end{figure}

In this section, we present the full analysis behind our results previously published in Ref.~\onlinecite{Fabrizi,*FabriziE}.
The zero field magnetic structure was determined at $T=15$~K in the incommensurate ferroelectric phase, in which the Mn atoms develop a cycloidal magnetic structure propagating along $\hat{\mathbf{b}}$. An essential component of the measurements was the poling electric field applied along $+\hat{\mathbf{c}}$ while cooling the sample inside the cryostat, enabling us to influence the magnetic domain population.
Fig.~\ref{Fig:structure} shows the final magnetic structure determined by performing circularly polarised X-ray non-resonant scattering experiments (see in Ref.\onlinecite{Fabrizi,*FabriziE}). An induced  weak magnetic contribution is also present at the Tb sites, with the same incommensurate periodicity $q_m=2\pi\tau \textbf{b}^*$ . These results have been determined by refining the polarisation dependence of scattered intensities for a complete star of wave-vectors (4,$\pm\tau$,$\pm1$), shown in Fig.~\ref{fig:H0stokes}. Given the inherent weakness of the non-resonant magnetic scattering signal (maximum count rate$=5$ counts per second\cite{noteCR}), a careful background subtraction is essential, with the result that 36 hours were required to measure the data shown in Fig.~\ref{fig:H0stokes}.

Here we shall derive these results and compare them with the prediction from the model established with unpolarised neutron scattering\cite{kenzelmann}. We will demonstrate that the model does not explain all the observed features and that a longitudinal magnetic component should be added to the Tb site. Using a semi quantitative description of the structure factor, we shall expand the arguments allowing for a description of the magnetic structure. A fit finally gives the exact parameters.

We start by labelling the different magnetic atom positions in the unit cell, as shown in Tab.~\ref{tab:positions}.

\begin{table}[!htb]
\caption{Mn and Tb atomic positions in the $Pbnm$ unit cell (space group n. 62). $\Delta^{Tb}_{a,b}$ describes the fractional atomic coordinate along $\hat{\mathbf{a}}$ and $\hat{\mathbf{b}}$ of the Wyckoff position (4c) occupied by the Tb atoms \cite{Blasco}.}
\label{tab:positions}
\begin{tabular}{|c|c|c|c|c|c|}
\hline\hline
atom&Wychoff&a&b&c&label\\
\hline
Mn & 4 b & 1/2  & 0 & 0 & M6 \\
&  & 0 &1/2 & 0 & M1 \\
&  &1/2 & 0 & 1/2 & M8 \\
&  & 0 & 1/2 &  1/2 & M3 \\
\hline
Tb & 4 c & 1-$\Delta_a^{Tb}$  & $\Delta_b^{Tb}$ &  1/4 & T1 \\
& & 1/2+$\Delta_a^{Tb}$ & 1/2+$\Delta_b^{Tb}$ &  1/4 & T6 \\
& & $\Delta_a^{Tb}$ & 1-$\Delta_b^{Tb}$ & 3/4 & T3 \\
& & 1/2-$\Delta_a^{Tb}$ & 1/2-$\Delta_b^{Tb}$ &  3/4 & T8 \\
\hline\hline
\end{tabular}
\end{table}

The symmetry splits the Tb moments into two orbits (T1, T6) and (T3, T8), however the moments, following Landau theory, are taken to be identical.
The magnetic structure deduced from neutron diffraction\cite{kenzelmann} is described according to:

\begin{equation}\label{eq:moments}
\begin{split}
&\mathbf{m}^{Mn}_{\Gamma3} = [0.0(5),3.9(1),0.0(7)]\mu_B\\
&\mathbf{m}^{Mn}_{\Gamma2} = [0.0(1),0.0(8),2.8(1)]\mu_B\\
&\mathbf{m}^{Tb}_{\Gamma3} = [0,0,0(1)]\mu_B\\
&\mathbf{m}^{Tb}_{\Gamma2} = [1.2(1),0(1),0]\mu_B\\
\end{split}
\end{equation}

\noindent where the numbers in brackets are error bars, and the absence of an error bar indicates that a moment component is forbidden by the cell's internal symmetry. The neutron experiment was however insensitive to the phases between several different moment components, including the phase difference between the Tb and Mn moments, and that between the $b$ and $c$ components of the Mn moment. With polarised neutrons\cite{yamasaki,*yamasakiE} and X-rays\cite{Fabrizi,*FabriziE} the latter was shown to be $\pm\pi/2$, such that the Mn moments form an elliptical cycloid. Taking the phase of the Mn moment component $m_b$ to be zero on site M1, then 
$\phi^{TM}_a$ is the phase of the Tb moments at T1 in the first orbit, whilst the phase difference $\phi^{TO}_a$ between the $m_a$ components at T1 and T3 defines the phases of the Tb moments in the second orbit.

We will now calculate the magnetic scattering amplitude based on Kenzelmann's model\cite{kenzelmann} for our measured reflections $\mathbf{k^\prime}-\mathbf{k}=\left(4, \alpha\tau, \beta\right)$, where $\alpha=\pm1$ selects the sign of the $k$ Miller index, $\beta=\pm1$, the sign of $l$, and the two different cycloidal domains are defined by $\gamma=\pm1$, where $\gamma=+1$ corresponds to an anticlockwise rotation when moving along $+\hat{\mathbf{b}}$ and looking from $+\hat{\mathbf{a}}$. To construct the structure factor for the spins $\mathbf s^{Mn}$, $\mathbf s^{Tb}$, we proceed as for the generic cycloid case given in the previous section adding in the moments on the Tb atoms obtaining:

\begin{equation}\label{eq:H0kenzSpin}
\begin{split}
&F_{NR}= \frac{-i\hbar\omega}{mc^2} \left\{2(s^{Mn}_b\hat{\mathbf{b}}-i\alpha\gamma s^{Mn}_c\hat{\mathbf{c}})\right.  \\
&\left. + \beta e^{i\alpha\phi^{TM}_a}\sin(8\pi\Delta_a^{Tb})(1+e^{i\alpha\phi^{TO}_a})s^{Tb}_a\hat{\mathbf{a}} \right\} \cdot\mathbf{P}_S
\end{split}
\end{equation}

To facilitate a qualitative comparison of the expected scattering arising from this model structure factor with the actual non-resonant X-ray magnetic scattering results\cite{Fabrizi,*FabriziE}, the above expression can be simplified by assuming that the scattering angle $2\theta=90^\circ$, that $\hat{\mathbf{b}}$ is perpendicular to the scattering plane and that the other two axes of the reference system, i.e. $\mathbf{k+k^\prime}$ and $\mathbf{k^\prime-k}$, correspond to the other two crystallographic axes. In writing  Eq.~\eqref{eq:H0kenzSpin}, we skipped all orbital terms since within this simplified but near real configuration,  $\mathbf{\hat a\cdot P}_L$ is zero, leaving for the Terbium, only the spin contribution $s _a^{Tb}$. Furthermore the Mn orbital moment is taken to be quenched so that we set $\mathbf m^{Mn}=2\mathbf s^{Mn}$. Then the scattering amplitudes for a circularly polarised beam incident, measured in the two linear polarisation channels $F_{\sigma^\prime}$ and $F_{\pi^\prime}$ are given by:

\begin{equation}\label{eq:H0kenz}
\begin{split}
F_{\sigma^\prime}=&M-i\epsilon\beta(\upsilon^\prime+i\alpha\upsilon^{\prime\prime})T_a\\
F_{\pi^\prime}=&i\epsilon M-\beta(\upsilon^\prime+i\alpha\upsilon^{\prime\prime})T_a
\end{split}
\end{equation}

\noindent where $\epsilon$ selects the handedness of the incident X-rays, and we use the shorthand notation:

\begin{equation}
\begin{split}\label{eq:simpH0ShHd1}
 M & =\frac{-i\hbar\omega}{mc^2}(2s_b^{Mn}-{\epsilon\alpha\gamma}{\sqrt{2}}s_c^{Mn})\\
 T_a & =\frac{-i\hbar\omega}{mc^2}\sin(8\pi\Delta_a^{Tb})s_a^{Tb}/\sqrt{2}\\
 \upsilon^\prime+i\alpha\upsilon^{\prime\prime} & =e^{i\alpha\phi^{TM}_a}(1+e^{i\alpha\phi^{TO}_a})
\end{split}
\end{equation}

If we consider only the Mn spins then $\left|F_{\sigma^\prime}\right|=\left|F_{\pi^\prime}\right|$, and so $P_1=0$, as in the case described in Eq.~\ref{eq:Poincare-Stokescycloid}. Also, as $F_{\sigma^\prime}$ would be real and $F_{\pi^\prime}$ imaginary, this means that $\left|F_{\sigma^\prime}+F_{\pi^\prime}\right|=\left|F_{\sigma^\prime}-F_{\pi^\prime}\right|$, such that $P_2=0$. Therefore, any departure from a circularly polarised diffracted beam is interpreted as arising from the Tb moments. We note that $P_2\propto|F_{\sigma^\prime}+F_{\pi^\prime}|^2-|F_{\sigma^\prime}-F_{\pi^\prime}|^2=-8\beta\upsilon^\prime MT_a$ whose sign is left unchanged by reversing the incident polarisation and the sign of $\tau$, while it is reversed by a change in $l$. Further inspection of the structure factors in Eqs.~\eqref{eq:H0kenz} and \eqref{eq:simpH0ShHd1} reveals that there will be an imbalance in the intensities between the twin reflections ($\pm\tau$) and for the opposite circular polarisations, where the intensity will be large for $\epsilon\alpha\gamma=-1$ and small for $\epsilon\alpha\gamma=+1$, but the handedness associated with the maximum satellite intensity is invariant with respect to changes in the sign of $l$.

We may test these predictions by inspection of Fig.~\ref{fig:H0stokes}, which reveals a clear imbalance in intensities, as expected from the Mn magnetic order.
In addition $I(\eta)$ is not independent of $\eta$, with extrema near $45^{\text o}$ and $135^{\text o}$ indicating that the scattered beam is not wholly circularly polarised and that $P_2$ is different from zero. However, the curves from this model do not fit the data so well, and the observed invariance of $P_2$ when the signs of incident polarisation, $l$ and $\tau$ are simultaneously reversed contradicts the model prediction.

To reproduce the values of $P_2$, the only solution found was the inclusion of a Tb magnetic moment component along $\hat{\mathbf{b}}$, a component to which the earlier neutron diffraction experiments\cite{kenzelmann} were largely insensitive. In addition we now take the orbital moments on the Tb ions into consideration,
with $l^{Tb}=s^{Tb}$ according to Hund's rules for the $^7F_6$ electronic configuration.

We can now construct the complete scattering amplitude:

\begin{widetext}
\begin{equation}
\begin{split}
&F_{NR}=\frac{-i\hbar\omega}{mc^2}\sum_s e^{2\pi i(hx_s+ky_s+lz_s)}\{\frac{1}{2}\mathbf{L}_s\mathbf{(K)\cdot P}_L+\mathbf{S}_s\mathbf{(K)\cdot P}_S\}\\
&=\frac{-i\hbar\omega}{mc^2}\left\{\left\{\!\beta e^{i\alpha\phi^{TM}_a}\!\sin(8\pi\Delta_a^{Tb})\left[(\!1+\!e^{i\alpha\phi^{TO}_a}) \frac{1}{2}l_a^{Tb}\hat{\mathbf{a}}\right]\!+\! i\beta e^{i\alpha\phi^{TM}_b}\!\cos(8\pi\Delta_a^{Tb})\left[(\!e^{i\alpha\phi^{TO}_b}-1) \frac{1}{2}l_b^{Tb}\hat{\mathbf{b}}\right]\right\}\cdot\mathbf{P}_L\right.\\
&\!+\!\left\{\left[2s_b^{Mn}\hat{\mathbf{b}}\!-\!{i\alpha\gamma}2s_c^{Mn}\hat{\mathbf{c}}\right]\!+\!
\beta e^{i\alpha\phi^{TM}_a}\!\sin(8\pi\Delta_a^{Tb})\!\left[(\!1\!+\!e^{i\alpha\phi^{TO}_a}) s_a^{Tb}\hat{\mathbf{a}}\right]
\!+\! \left. i\beta e^{i\alpha\phi^{TM}_b}\!\cos(8\pi\Delta_a^{Tb})\!\left[(e^{i\alpha\phi^{TO}_b}\!-1\!)s_b^{Tb}\hat{\mathbf{b}}\right]\right\}\cdot\mathbf{P}_S\right\}
\end{split}
\end{equation}
\end{widetext}

$\phi^{TM}_b$ and $\phi^{TO}_b$ being defined in the same way as $\phi^{TM}_a$ and $\phi^{TO}_a$, the Terbium magnetic structure has four phase parameters but their freedom may be restricted. The high magnetocrystalline anisotropy of the Terbium favours a moment with a fixed direction rather than rotating in a cycloid, as confirmed by observations. For a high magnetic field applied along $\mathbf{b}$\cite{AliouanePRL}, the Mn moments have a cycloidal arrangement in the $\mathbf{a-b}$ plane, while the Tb moments have a fixed oblique direction in the same plane. Whilst below $7$~K, Quezel \emph{et al.}\cite{Quezel} also found that the Tb moments lie along two symmetrical oblique directions. This implies that $\phi^{TM}_a$ and $\phi^{TM}_b$ as $\phi^{TO}_a$ and $\phi^{TO}_b$ differ only by $0$ or $\pi$, since any other outphasing between $\mathbf{a}$ and $\mathbf{b}$ components would produce a cycloidal rotation.
Some arguments developed in Appendix I show that to obtain the correlation shown in Fig.~\ref{fig:H0stokes} between the sign of $P_2$ on one side, and the signs of $k$, $l$ and the incident polarisation state on the other, we must have $\phi^{TO}_a=\phi^{TO}_b=\pi$, while $\phi^{TM}_b=0, \pi$ is preferred. Our experiment is insensitive to the $\mathbf{a}$ Tb moment component, as well as to $\phi^{TM}_a$.
When the full calculation is performed, taking the true geometry into account, the best fit to the data shown by the solid lines in Fig.~\ref{fig:H0stokes}, is obtained for a total Tb magnetic moment component along $\hat{\mathbf{b}}$ of $m_b^{Tb}=1.0\pm0.3\mu_B$, with a phase shift between the two Tb orbits of $\phi^{TO}_b=(1.0\pm0.2)\pi$ and a phase difference $\phi^{TM}_b=(0.0\pm0.1)\pi$ between one Mn atom and the subsequent Tb atom moving along $\hat{\mathbf{c}}$. The domain population is found to be $83(2)\%$ of the cycloidal domain in which the transverse spiral of the Mn atoms is clockwise, when moving along $+\hat{\mathbf{b}}$ and looking from $+\hat{\mathbf{a}}$.

The magnetic structure is not yet completely determined since the phase difference $\phi^{TM}_a$ can be $0$ or $\pi$, leaving the sign of the Tb $\mathbf{a}$ component unknown. In the ${\mathbf{a-b}}$ plane the Tb moment points to a direction near $45^{\text o}$ from the axes but in which quadrant remains unspecified. In the ideal cubic perovskite structure, all four quadrants are equivalent, while in the real arrangement they are not. Xiang {\it et al}~\cite{Xiang} predicted from theory an angle of $145$ degrees between $\mathbf{a}$ and the Tb moment labelled T1 in Table 1.

%
%

\section{$\mathrm{TbMnO_3}$ 0$<$H$<$H$_C$}\label{s:Hn0}

Neutron diffraction\cite{AliouanePRL} and X-ray resonant scattering\cite{Aliouane06,Strempfer,WalkerPhysB} measurements have found no evidence of a change in the magnetic structure in applied magnetic fields less than the critical field $H_C$ for the ferroelectric polarisation flop. However, in addition to the charge reflections observed in X-ray experiments at the double harmonic positions $\tau_L=2\tau$, which arise from a quadratic magnetoelastic coupling between the spins and the lattice, in low applied magnetic fields charge reflections were also observed at the single harmonic positions\cite{Aliouane06,Strempfer}, which were qualitatively interpreted in terms of a linear magnetoelastic coupling. Therefore rather than jumping straight to the $H>H_C$ phase, it was decided to investigate the effect of low applied magnetic fields first.

\begin{figure}
\centering
\includegraphics[width=.425\textwidth]{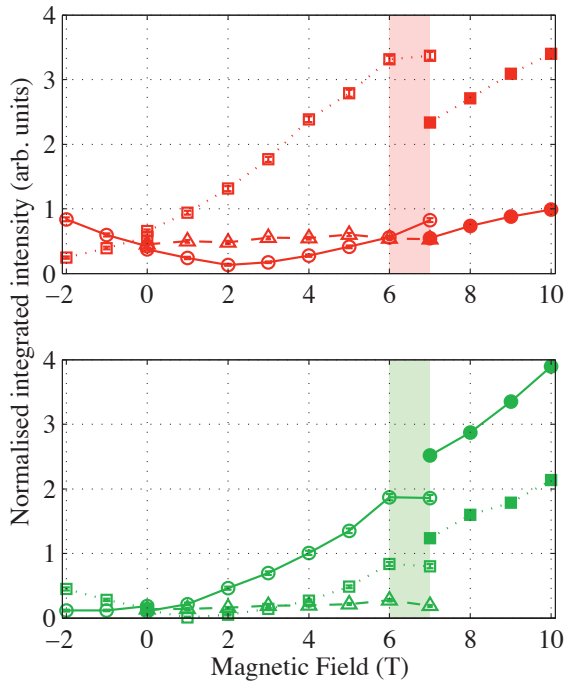}
\caption{Variation of the scattering from the (4 $\tau$ -1) reflection in TbMnO$_3$ at $T=12$~K as a function of magnetic field for circular left (top) and circular right (bottom) X-rays incident. Measurements were performed with the analyser at $\eta=45^\circ$ ($\circ$), $90^\circ$ ($\triangle$) and $135^\circ$ ($\Box$). At $H=7$~T two peaks are seen, one at the incommensurate position (open symbols) and the other at $\tau=\frac{1}{4}$ (full symbols). The intensities for $\tau=\frac{1}{4}$ have been divided by three to simplify combining the two data sets in the same figure.
}\label{fieldcurves}
\end{figure}

\begin{figure}
\centering
\includegraphics[width=.425\textwidth]{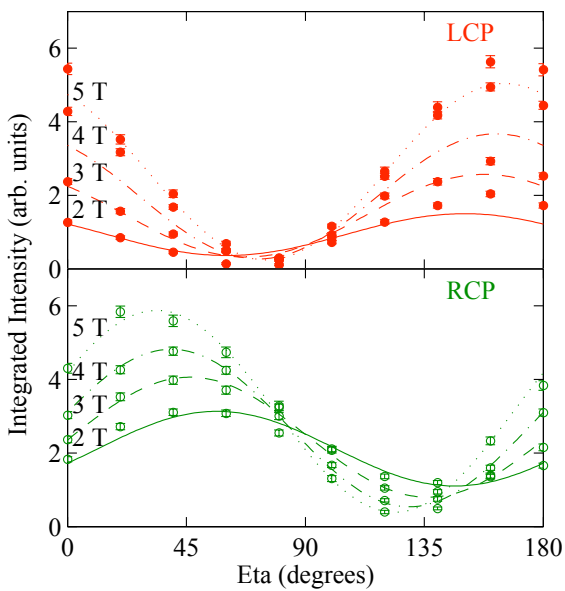}
\caption{\label{fig:magstokes}Stokes dependence of scattering from the reflection (4,$\tau$,$-1$) at $T=15$~K for $6.85$~keV LCP and RCP X-rays incident on TbMnO$_3$. The lines are fits to equation~\eqref{eq:finterf}. 
}
\end{figure}

In this section, we present a complete analysis of our results published in Ref.~\onlinecite{WalkerScience}.
The experiments were performed using the same experimental setup as for the zero field measurements, apart from the fact that the electric stick was inserted into the 10 T Oxford Instruments cryomagnet rather than an orange cryostat. The sample was cooled into the multiferroic phase ($T<T_N=28$~K) whilst applying an electric field along $+\hat{\mathbf{c}}$. Reciprocal space scans along $k$ over the reflection (4 $\tau$ -1) were measured for $6.16$~keV LCP and RCP X-rays incident, with the LiF (220) analyser at $\eta=45^\circ$, $90^\circ$ and $135^\circ$, as a function of increasing applied magnetic field (see Figure~\ref{fieldcurves}). Below $H\simeq6$~T, no change was seen in the ordering wave-vector, and  the intensities measured for $\eta=90^\circ$ are approximately invariant with increasing applied magnetic field, indicating no change in the cycloidal domain populations. However, the scattered intensity for $\eta=45^\circ$ and $\eta=135^\circ$ is essentially quadratic in the applied field. In another experiment, performed using a Cu (220) analyser at an incident X-ray energy of $6.85$~keV, we measured the variation in the Stokes dependence as a function of the applied magnetic field, after poling in an electric field along $-\hat{\mathbf{c}}$ (Fig.~\ref{fig:magstokes}). This demonstrates that with increasing the magnetic field the scattering becomes more linearly polarised.

Such a variation cannot be explained by modifications to the magnetic structure. Instead, this behaviour is due to the magnetostriction induced ionic displacements resulting in charge scattering at the magnetic ordering wave-vector\cite{WalkerScience}. The resultant Thomson scattering amplitude varies linearly as a function of field, and the 
combination of the unchanging magnetic scattering and this magnetic field dependent charge scattering produces the parabolic intensity variation.

Indeed let us look at the graph Fig.~\ref{fieldcurves}, showing the intensities at $\eta=45^\circ$ and $\eta=135^\circ$. Their difference seems to vary proportionally to the field between -2T and +2 T at least, and the sign of this variation is opposite from one of the incident circular polarization to the other. On Fig.~\ref{fig:magstokes} we see that the difference of intensity at $\eta=0^\circ$ and $\eta=90^\circ$ increases with the field, though at a different rate for the two opposite circular polarizations.
Those qualitative features can be simulated using a scattering amplitude which combines the non-resonant magnetic scattering with charge scattering:
\begin{equation}\label{eq:finterf}
F=(A_C+iB_C)+F_{NR},
\end{equation}
where $A_C+iB_C$ is proportional to the scattering amplitude arising from a lattice distortion $\delta\mathbf{r}_s$: $\sum_s\exp(i\mathbf{K}\cdot[\mathbf{r}_s+\delta\mathbf{r}_s])f_s(K,\omega)$.

For a qualitative interpretation we may neglect the small Tb moment keeping only the Mn and use the same simplifications as for Eq.~\eqref{eq:H0kenz}. Within these assumptions $2\theta\simeq90^\circ$, such that the Thomson term will appear only in $\sigma-\sigma^\prime$, and we obtain:
\begin{equation}
\begin{split}
F_{\sigma^\prime}&=M+A_C+iB_C,\\
F_{\pi^\prime}&=i\epsilon M,
\end{split}
\end{equation}
$M$, defined in Eq.~\eqref{eq:simpH0ShHd1} is pure imaginary; when changing the sign $\epsilon$ of the incident polarization, it changes its magnitude, smaller in circular right, while keeping the same sign. Then

\begin{equation}
\begin{split}
\left|F_{\sigma^\prime}\right|^2-\left|F_{\pi^\prime}\right|^2&=
-2iM B_C+B_C^2+A_C^2\\
\left|F_{\sigma^\prime}+F_{\pi^\prime}\right|^2-
\left|F_{\sigma^\prime}-F_{\pi^\prime}\right|^2&=4i\epsilon M A_C
\end{split}
\end{equation}

\noindent The first and second line explain the features observed Fig.~\ref{fig:magstokes} and~\ref{fieldcurves} respectively, if assumed that $A_C$ and $B_C$ are proportional to the field, allowing a simple semi-quantitative determination of their values.

\begin{table}[h]
\caption{Measured real and imaginary part $A_C$ and $B_C$, of the displacement Thomson
structure factor for one cell or 4 formula units, in units of $r_e$. The crystal was initially poled in zero magnetic field with an electric field along $-\mathbf c$, opposite to the case discussed in Sect. \ref{s:H0}.
}
\setlength{\unitlength}{1mm}
\newcommand{\bloc}[1]{\makebox(12,3.3)[b]{#1}}
\newcommand{\bloo}[1]{\makebox(20,3.3)[b]{#1}}
%
\begin{tabular}
{|c|c|c|c|}
\hline\hline
\bloc{$E$ (keV)}&\bloc{$h\,k\,l$}&\bloo{$A_C/H$ ($r_e$/T)}&\bloo{$B_C/H$ ($r_e$/T)} \\
\end{tabular}
\vskip -0.5mm
\begin{tabular}
{|c|}
\hline
\makebox(68.6,3.4)[b]{$H<6$ T (Sect. \ref{s:Hn0})} \\
\end{tabular}
\vskip -.5mm
\begin{tabular}
{|c|c|c|c|}
\hline
\bloc{$6.16$}&\bloc{$4\,\tau\,1$}&\bloo{$-0.0068(2)$}&\bloo{$-0.0011(3)$} \\
&\bloc{$4\,\tau\,\overline 1$}&\bloo{$-0.0064(2)$}&\bloo{$\ \;\,0.0039(3)$} \\
\bloc{$6.85$}&\bloc{$4\,\tau\,1$}&\bloo{$-0.0065(2)$}&\bloo{$-0.0018(3)$} \\
&\bloc{$4\,\tau\,\overline 1$}&\bloo{$-0.0065(2)$}&\bloo{$\ \;\,0.0040(3)$} \\
\bloc{$7.77$}&\bloc{$4\,\tau\,1$}&\bloo{$-0.0051(2)$}&\bloo{\ \;\,%
$0.0004(3)$} \\
&\bloc{$4\,\tau\,\overline 1$}&\bloo{$-0.0050(2)$}&\bloo{$\ \;\,0.0051(3)$} \\
\end{tabular}
\vskip -0.5mm
\begin{tabular}
{|c|}
\hline
\makebox(68.6,3.4)[b]{$H= 10$ T (Sect. \ref{s:H10})} \\
\end{tabular}
\vskip -0.5mm
\begin{tabular}
{|c|c|c|c|}
\hline
\bloc{$6.16$}&\bloc{$4\,\overline \tau\,\overline 1$}&%
\bloo{$-0.0060(4)$}&\bloo{$-0.0024(4)$} \\
\hline\hline
\end{tabular}
\vskip -3.2mm
\label{tab:ExpStrucFac}
\end{table}

More accurately we measured the Stokes dependence of different reflections as a function of magnetic field and, fitting the data with exact formulae, extracted the Thomson scattering amplitude, shown in Table \ref{tab:ExpStrucFac}.
Then by exploiting the variation in the atomic dispersion corrections for Mn and Tb at different non-resonant incident energies we obtained a quantitative estimate of the different off-centre ionic displacements\cite{WalkerScience}. 

\begin{figure*}
\includegraphics[width=0.85\textwidth]{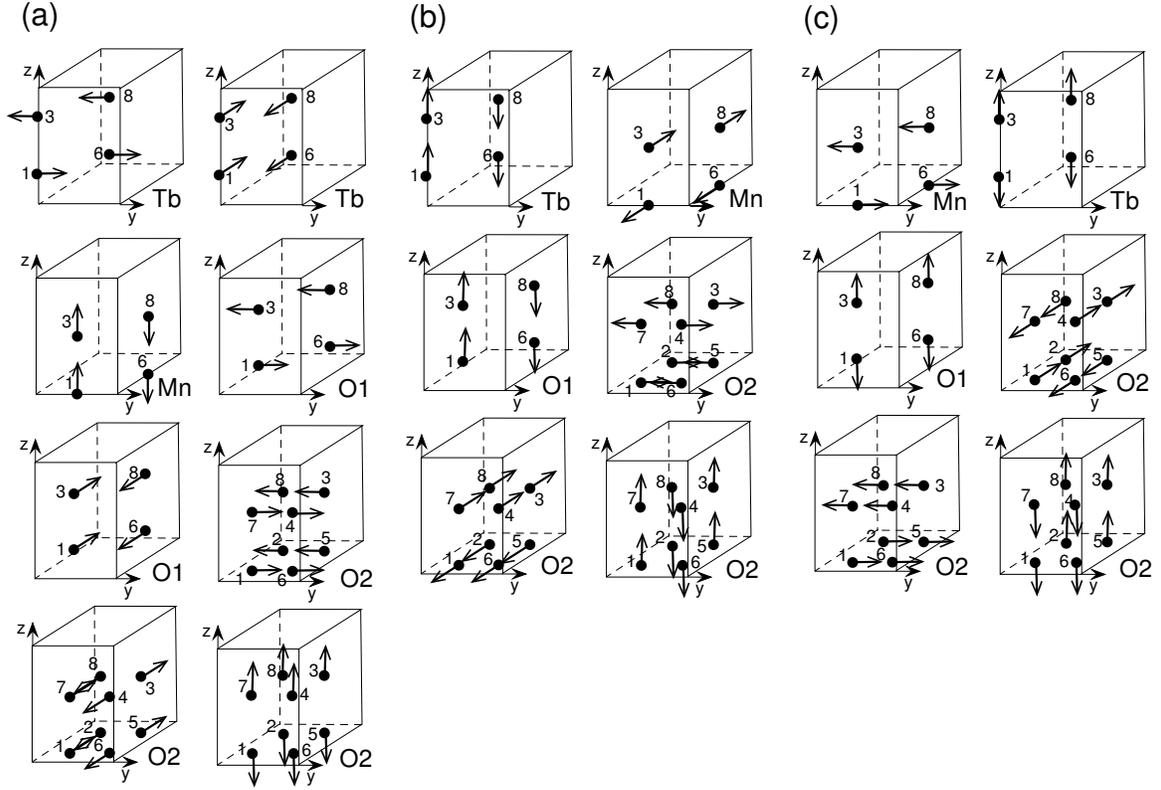}
\centering
\caption{\label{fig:modes}The different displacement modes in (a) $\Gamma_1$, (b) $\Gamma_2$ and (c) $\Gamma_4$ visible at (4 $\pm\tau$ $\pm1$) as listed in Table~\ref{tab:modes}.}
\end{figure*}


Here below, and in Appendix II, we present some arguments that were not fully developed in our previous paper\cite{WalkerScience}, showing how the magnetostrictive displacements are represented in modes with specific symmetry properties and some phase defined with respect to the magnetic structure. We also discuss the insight which the results provide into the origin of the ferroelectric polarisation.

The displacement modes are linear combinations of the atomic displacements, which can be separated into two classes, since the magnetic structure splits the eight positions in the $Pbnm$ spacegroup (see Appendix II) into two independent orbits:
\begin{equation}
\begin{array}{rlrl}
\Delta_{\alpha1}&\!=\!\delta_1\!+\!\delta_3\!+\!\delta_6\!+\!\delta_8\,&\Delta_{\alpha2}&\!=\!\delta_5\!+\!\delta_7\!+\!\delta_2\!+\!\delta_4\\
\Delta_{\beta1} &\!=\!\delta_1\!-\!\delta_3\!+\!\delta_6\!-\!\delta_8\,&\Delta_{\beta2} &\!=\!\delta_5\!-\!\delta_7\!+\!\delta_2\!-\!\delta_4\\
\Delta_{\gamma1}&\!=\!\delta_1\!+\!\delta_3\!-\!\delta_6\!-\!\delta_8\,&\Delta_{\gamma2}&\!=\!\delta_5\!+\!\delta_7\!-\!\delta_2\!-\!\delta_4\\
\Delta_{\delta1}&\!=\!\delta_1\!-\!\delta_3\!-\!\delta_6\!+\!\delta_8\,&\Delta_{\delta2}&\!=\!\delta_5\!-\!\delta_7\!-\!\delta_2\!+\!\delta_4.\\
\end{array}
\end{equation}
They may be rewritten according to
\begin{equation}
\Delta_{\alpha\pm}=\frac{1}{2}(\Delta_{\alpha1}\pm\Delta_{\alpha2}).
\end{equation}
In order to describe the symmetry of the modes let us consider the actions of the generators of the little group $G_k$:\\
\begin{center}
\begin{tabular}{c|cccc}
&1&$2_y$&$m_{xy}$&$b$\\
\hline
$\Gamma_1$&1&1&1&1\\
$\Gamma_2$&1&1&-1&-1\\
$\Gamma_3$&1&-1&1&-1\\
$\Gamma_4$&1&-1&-1&1\\
\end{tabular}\\
\end{center}
i.e. the propagation wave-vector preserving subgroup of $Pbnm$. 
How each mode is assigned to one specific irrep, its phase and whether it is visible at the measured peaks ($4$, $\pm\tau$, $\pm1$) is described in Appendix II, while these properties are summarised in Table~\ref{tab:modes}.

\begin{table}[!htb]
\caption{The status of all the displacement modes in both low (section~\ref{s:Hn0}) and high (section~\ref{s:H10}) magnetic field phases. For each component $a$, $b$ or $c$ of the modes $\Delta$, the irrep to which it belong, the phase relative to the magnetic component $M_b^{Mn}$ -- \emph{R}eal or \emph{I}maginary, and whether they are visible or extincted for the A-type peak in the experiment are given. The sites column indicates which sites have the visible mode in their structure factor. Note: none of the magnetic components produces any displacement in $\Gamma_3$, but they are included here for completeness.}\label{tab:modes}
\begin{tabular}{|c|ccc|ccc|ccc|ccc|}
\hline\hline
mode&&a&&&b&&&c&&&sites&\\
\hline
$\Delta_{\alpha+}$&$\Gamma_3$&   &ext.&$\Gamma_1$&$R$&ext.&$\Gamma_4$&$I$&ext.&&&\\
$\Delta_{\alpha-}$&$\Gamma_2$&$I$&ext.&$\Gamma_4$&$R$&ext.&$\Gamma_1$&$I$&ext.&&&\\
$\Delta_{\beta+} $&$\Gamma_2$&$R$&vis.&$\Gamma_4$&$I$&vis.&$\Gamma_1$&$R$&vis.&O$_2$&&Mn\\
$\Delta_{\beta-} $&$\Gamma_3$&   &vis.&$\Gamma_1$&$I$&vis.&$\Gamma_4$&$R$&vis.&O$_2$&O$_1$&Tb\\
$\Delta_{\gamma+}$&$\Gamma_1$&$R$&vis.&$\Gamma_3$&   &vis.&$\Gamma_2$&$R$&vis.&O$_2$&O$_1$&Tb\\
$\Delta_{\gamma-}$&$\Gamma_4$&$R$&vis.&$\Gamma_2$&$I$&vis.&$\Gamma_3$&   &vis.&O$_2$&&\\
$\Delta_{\delta+}$&$\Gamma_4$&$I$&ext.&$\Gamma_2$&$R$&ext.&$\Gamma_3$&   &ext.&&&\\
$\Delta_{\delta-}$&$\Gamma_1$&$I$&ext.&$\Gamma_3$&   &ext.&$\Gamma_2$&$I$&ext.&&&\\
\hline\hline
\end{tabular}
\end{table}

Since the magnetic structure of TbMnO$_3$ is described using the $\Gamma_2$ and $\Gamma_3$ irreps, whilst the magnetic field induced moment belongs to $\Gamma_2$, the interaction between the two will give rise to two classes of displacements: $\Gamma_1=\Gamma_2\otimes\Gamma_2$ and $\Gamma_4=\Gamma_3\otimes\Gamma_2$, which are shown in Figure~\ref{fig:modes}. The displacement structure factor is even in $\pm h$ and $\pm k$, but whilst it is even in $\pm l$ for the $\Gamma_4$ part, it is odd for the $\Gamma_1$ part. Therefore, in order to identify the particular atomic displacement modes giving rise to the charge scattering, the Stokes dependence of the scattering was measured for two reflections ($4,\tau,\pm1$). The $\Gamma_1$ class of modes is complicated since it arises due to interactions, symmetric and antisymmetric, between numerous moments. However, the $\Gamma_4$ class arises due to only the symmetric interaction between the induced moment and $S_b^{Tb}$, making it more tractable. The different extracted modes are listed in Table~\ref{tab:disps}, including the displacement of the Tb ions along $\hat{\mathbf{c}}$, of maximum magnitude $-19\pm2$~fm/T.

\begin{table}[htb]
\caption{Displacements of the Mn, Tb and oxygen ions along the $\hat{\mathbf{a}}$, $\hat{\mathbf{b}}$ and $\hat{\mathbf{c}}$ axes in femtometres per Tesla in the low field phase.}\label{tab:disps}
\begin{tabular}{|r|p{.5cm}lcl|}
\hline\hline
&&&&\\
$\Gamma_1$  & $i\delta_b^{Tb}$&$+6.8\delta_a^{Tb}$ & $=$ & $-6\pm6$\\[1ex]
            & $\delta_c^{Mn}$& & $=$ & $-4\pm4$\\[1ex]
            & $\delta_c^{O2}$&$-2.4\delta_a^{O1}-0.34i\delta_b^{O1}$ & &\\
            & &$+3.2\delta_a^{O2}+0.12i\delta_b^{O2}$ & $=$ & $-50\pm20$\\[1ex]
\hline
&&&&\\
$\Gamma_4$  & $\delta_c^{Tb}$& & $=$ & $-19\pm2$\\[1ex]
            & $i\delta_b^{Mn}$& & $=$ & $+5\pm5$\\[1ex]
            & $\delta_c^{O1}$&$-10\delta_a^{O2}-0.37i\delta_b^{O2}$ & &\\
            & &$-0.36\delta_c^{O2}$ & $=$ & $-72\pm4$\\[1ex]
\hline\hline
\end{tabular}
\end{table}

We now get an insight into the onset of ferroelectricity in TbMnO$_3$. These Tb displacements are in anti-phase along $\hat{\mathbf{c}}$, and will therefore sum to give zero net ferroelectric polarisation, consistent with the plateau seen in the measured bulk polarisation\cite{KimuraTMO} for $H<H_C$. They result from a symmetric exchange striction between uniform induced $\Delta m_b^{Tb}$ and modulated native $m_b^{Mn}$ moments which are in anti-phase moving along $\hat{\mathbf{c}}$. In the zero field structure, $m^{Tb}_b$ are also in antiphase moving along $\hat{\mathbf{c}}$, such that the same interaction with $m^{Mn}_b$, gives rise to in-phase displacements of the Tb ions along $\hat{\mathbf{c}},$ and these contribute to the spontaneous ferroelectric polarisation (see Fig.~4 in Ref.~\onlinecite{WalkerScience}).
Let us suppose that a given force produces the same displacement, whether magnetostrictive or spontaneous.
Then, by combining our knowledge regarding the magnetic field induced ionic displacements, the magnetisation as a function of applied magnetic field\cite{KimuraTMO}, and the magnitude of the Tb moment in zero magnetic field\cite{Fabrizi,*FabriziE}, we can estimate the zero-field ionic displacements. An induced moment equivalent to that of the zero field moment $m^{Tb}_b=1\mu_B$ occurs for an applied field of $2.2$~T, and with a factor $1/2$ to account for comparison of the maximum of a sine to the average of a sine square, we estimate the zero-field displacement of the Tb ions along $\hat{\mathbf{c}}$ to be of average amplitude $-21\pm3$~fm. This results in a net Tb contribution to the spontaneous ferroelectric polarisation of $175\pm25$~$\mu$Cm$^{-2}$ from ionic displacements, representing one quarter of the total polarisation measured\cite{KimuraTMO}. Yet this should be taken as just an order of magnitude, because assuming that the Tb displacement depends only on the exchange striction force is not quite correct. Indeed a complete description should include the displacements of oxygen ions, which behave differently in both cases. Also we should remember that this is a secondary effect, occurring once the ferroelectric displacement has been trigged by some antisymmetric, such as Dzyaloshinskii-Moriya, interaction. Even though being not precisely quantitative, our result brings in a strong argument in favour of a displacive\cite{Sergienko,Malashevich,Xiang} rather than electronic mechanism to drive the ferroelectricity in type II magnetoelectric multiferroics as exemplified by TbMnO$_3$.

%
%

\section{$\mathrm{TbMnO_3}$ H$>$H$_C$}\label{s:H10}

\begin{figure}[b]
\centering
\includegraphics[width=0.425\textwidth]{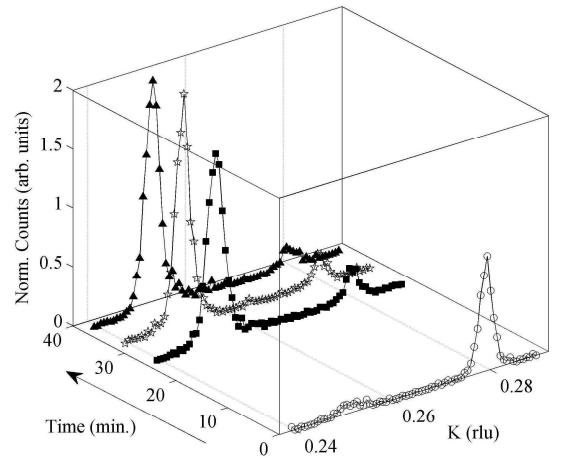}
\caption{Evolution of the co-existing commensurate and incommensurate reflections (4 $\tau$ -1) in TbMnO$_3$ as a function of time in an applied magnetic field $H=7$~T at $T=12$~K measured with LCP X-rays incident and the analyser at $\eta=135^\circ$. 
}\label{fig:time}
\end{figure}

Striking evidence for a strong magnetoelectric coupling in TbMnO$_3$ comes from the observation of a flopping of the ferroelectric polarisation from the $\hat{\mathbf{c}}$ to the $\hat{\mathbf{a}}$ axis on application of sufficiently large ($\sim6$~T) magnetic fields along the $\hat{\mathbf{a}}$ or $\hat{\mathbf{b}}$ axes\cite{KimuraTMO}. Therefore, it is no surprise that this transition has been studied in great depth by numerous different probes\cite{Arima05,Aliouane06,AliouanePRL,Strempfer,WalkerPhysB,Barath}. Initial scattering studies focussed on the ordering wave-vector, revealing that the polarisation flop transition is concomitant with a transition from an incommensurate to a commensurate magnetic structure with $\mathbf{k}_m=(0 \frac{1}{4} 0)$\cite{Arima05,Aliouane06}, opening up the possibility of the ferroelectric polarisation arising due to exchange striction. However, Aliouane \emph{et al.}, using neutron diffraction, revealed that the flopping of the ferroelectric polarisation from the $\hat{\mathbf{c}}$ to the $\hat{\mathbf{a}}$ axis in the high field commensurate phase is accompanied by the flopping of the Mn spin cycloid from the $\mathbf{b-c}$ to the $\mathbf{a-b}$ plane\cite{AliouanePRL}. The high-field magnetic structure can therefore be described using the $\Gamma_1$ and $\Gamma_3$ irreps according to $m_1=(2.83,0.51,0)\,\mu_B$/Mn and $m_3=(0.55,3.79,0)\,\mu_B$/Mn, with a phase difference between the two irreps of $0.474\pi$. The Tb magnetic structure Tb is also modified, from an incommensurate transverse and longitudinal sinusoidal wave with components along the $\hat{\mathbf{a}}$ and $\hat{\mathbf{b}}$ axes respectively\cite{Fabrizi,*FabriziE}, to an ordering commensurate with the underlying crystal lattice, i.e. $\tau_\mathrm{Tb}=0$, with a total magnetic moment of $7.24\,\mu_B$ composed of a $6.07\,\mu_B$ antiferromagnetic component along $\hat{\mathbf{a}}$ and a $3.92\,\mu_B$ ferromagnetic component along $\hat{\mathbf{b}}$\cite{AliouanePRL}. Although, it is worth noting that X-ray resonant scattering measurements performed in this commensurate high field phase revealed scattering at $\tau=0.25$ at the Tb $L_{III}$ edge, implying some ordering of the Tb with this wave-vector\cite{WalkerPhysB}. It is therefore likely that the Mn ordering with wave-vector $\tau=0.25$ induces a small moment on the Tb with the same wave-vector, but the magnitude of this moment is insignificant in comparison with the $\tau_\mathrm{Tb}=0$ Tb moment. Given the strong interest in this phase, it was therefore fitting to complete our survey of the magnetic field phase diagram of TbMnO$_3$ using circularly polarised X-ray non-resonant scattering.

\begin{figure}
\centering
\includegraphics[width=.475\textwidth]{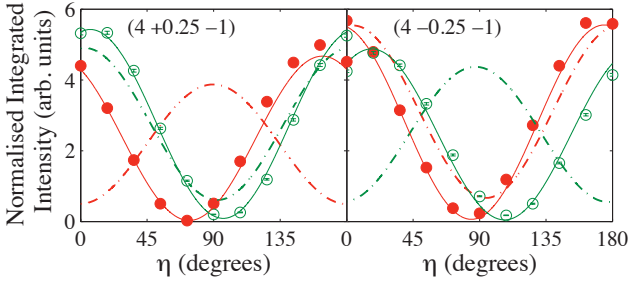}
\caption{Stokes dependence of the scattering from the (4 $\pm0.25$ -1) reflection in TbMnO$_3$ at $T=14$~K and $H=10$~T, for LCP (\textcolor[rgb]{1,0.00,0.00}{$\bullet$}) and RCP (\textcolor[rgb]{0.00,0.7,0.00}{$\circ$}) X-rays incident, compared with two models: magnetic structure proposed from neutron diffraction (dot-dash line), and the magnetic structure combined with ionic displacements (solid line), where the Thomson component is identical for the two reflections.
}\label{fig:stokesH10}
\end{figure}

If we now return to the previous experiment, Figure~\ref{fieldcurves}, then it is clear that the parabolic variation in $I(\eta=45^\circ,135^\circ)$ in low applied magnetic fields stops abruptly above $6$~T, and then there is a jump in the intensity for $k=0.25$. Whilst at $H=7$~T the $k$-scans reveal two reflections (Fig.~\ref{fig:time}), at the incommensurate position $k=0.28$ and at the commensurate position $k=0.25$. This apparent phase coexistence has been previously observed in Raman scattering, and might be consistent with the development of fluctuating commensurate and incommensurate domains contributing to this discontinuous magnetic transition\cite{Barath}. On performing the $k$-scan immediately after increasing the magnetic field to $7$~T, it was found that the incommensurate reflection was more intense than the commensurate one. However, when the scan was repeated twenty minutes later, it was found that the commensurate peak was now more intense, and that as a function of time, the incommensurate peak continued to get smaller (see Fig.~\ref{fig:time}).

In another experiment the Stokes dependence was measured at $T=14$~K and $H=10$~T, in the high field phase at the (4 $\pm\tau$ -1) reflections for LCP and RCP incident, see Fig.~\ref{fig:stokesH10}. Before applying the magnetic field, the crystal was poled in an electric field along $-\hat{\mathbf{c}}$. At the end of the measurement, the magnetic field was lowered back to zero then the sense of the ferroelectric-cycloidal domain was checked. In order to obtain a quantitative understanding of the measurements shown in Fig.~\ref{fig:stokesH10}, we construct the amplitude at (4 $\pm0.25$ -1) for the neutron diffraction determined magnetic structure\cite{AliouanePRL}, simplified as a Mn elliptical cycloid with main axes $\hat{\mathbf{a}}$ and $\hat{\mathbf{b}}$; Tb is not added since $\tau_{Tb}=0$:

\begin{equation}
\frac{mc^2}{-i\hbar\omega}f_{NR}=\left(2(m^{Mn}_b\hat{\mathbf{b}}-i\alpha\gamma m^{Mn}_a\hat{\mathbf{a}})\right)\cdot\mathbf{P}_S.
\end{equation}
Setting $M_a=\frac{\hbar\omega}{mc^2}\sqrt 2\alpha\gamma s^{Mn}_a$, $M_b=\frac{\hbar\omega}{mc^2}2 s^{Mn}_b$ and going on with the same simplified geometry as in previous sections, we obtain the scattering amplitudes, including the Thomson terms

\begin{equation}
\begin{split}\label{eq:H10simp}
F_{\sigma^\prime}&=-iM_b+i\epsilon M_a+A_C+i B_C,\\
F_{\pi^\prime}&=\epsilon (M_b +\epsilon M_a),
\end{split}
\end{equation}
giving

\begin{equation}
\begin{split}\label{eq:H10P1P2}
&\left|F_{\sigma^\prime}\right|^2-\left|F_{\pi^\prime}\right|^2=A_C^2 + B_C^2 -2(M_b-\epsilon M_a) B_C \\
&\makebox[25mm]{}- 4\epsilon M_a M_b,\\
&\left|F_{\sigma^\prime}+F_{\pi^\prime}\right|^2-\left|F_{\sigma^\prime}-F_{\pi^\prime}\right|^2=4(\epsilon M_b+M_a)A_C.
\end{split}
\end{equation}

This is to be compared with the data in Fig.~\ref{fig:stokesH10}. The figure includes the simulations made within the exact neutron model and the true geometry, with (solid lines) and without (dot-dash lines) the Thomson scattering. Without any Thomson scattering a large $P_1$ is obtained, whose sign reverts together with the incident polarisation, while the experiment shows a large $P_1$, keeping the same sign in all cases, as predicted by the first of Eqs.~\eqref{eq:H10P1P2}, if the Thomson $A_C$, $B_C$ terms are dominant. Though smaller, the magnetic terms $M_a$, $M_b$ are sufficient to produce some difference between intensities in $\sigma^\prime$ at both incident polarisation, through a difference in $(M_b-\epsilon M_a)$. Furthermore the shift of the minimum away from $\eta=90^\circ$, revealing a non zero $P_2$, is explained by the second of Eqs.~\eqref{eq:H10P1P2}: since it is known that $\left|M_b\right|>\left|M_a\right|$\cite{AliouanePRL}, we would expect the sign of $P_2$ to be reversed for LCP vs RCP incident, and since $M_a\neq0$ the absolute value of $P_2$ will also be changed, which is consistent with the asymmetric displacement of the minima. These contributions of magnetic terms, otherwise known from neutron diffraction, allow finding the sign and scale of Thomson terms through a fitting procedure. Moreover, when two fits are made independently for both (4 $\pm$0.25 -1) reflections, the same Thomson term is found, indicating that, as in the low field case, the Thomson scattering term is invariant under $\alpha$, the sign of $\mathbf q$. The fit is consistent for one particular sign of the magnetic cycloid only, which is such determined, see the discussion below. In order to determine which displacement modes are active within the high field phase it is again necessary to perform a symmetry analysis.

The parts of the magnetisation with a zero propagation wave-vector are the ferromagnetic $b$ and antiferromagnetic $a$ components of the Tb moments which both belong to irrep $\Gamma_2$. Therefore, since the Mn magnetic modulations belongs to $\Gamma_1$ and $\Gamma_3$\cite{AliouanePRL}, the displacement modes arising through magnetostriction will belong to $\Gamma_4=\Gamma_2\otimes\Gamma_3$, as in the phase $H<H_C$, and $\Gamma_2=\Gamma_2\otimes\Gamma_1$, which are listed in Table~\ref{tab:modes} and shown in Figure~\ref{fig:modes}. Whilst the same arguments may be used to assign representations to the modes and determine the extinction conditions, the relative phase of the magnetic and lattice modulations happens to be less well determined, with the magnetic structure now locked to the lattice. However, 
assuming that the locking interaction is a second order effect,
we still used the same method as for the incommensurate structure. The phases, given in Table~\ref{tab:modes}, are approximate, valid inasmuch as some $n$ glide of the crystal structure applies to the magnetic structure.

Regrettably, with data only for $l=-1$ at a single energy, this is insufficient to be able to determine the individual ionic displacements, as was done in the low field phase. However, they still yield some interesting information. First we observe that both ferroelectric phases have one feature in common, the Mn magnetisation along $\hat{\mathbf{b}}$, which belongs to the irrep $\Gamma_3$, and hence induces displacements in $\Gamma_4$ as discussed above. At $6.16$~keV, the real part of the Thomson amplitude ($A_C$ from Eqn.~\eqref{eq:finterf}) is expected to come mainly from that common component, since the other displacements contribute to $A_C$ via only a weak atomic dispersion factor. Below $H_C$, at $6.16$~keV, the fits to the Stokes data give $A_C/H=-0.0066$ per Tesla per unit cell, i.e. per four formula units. Given an estimate of $0.45 \mu_B$~/T for the Tb susceptibility, that equates to an $A_C$ of $-0.0146$ per $\mu_B$ on Tb along $\hat{\mathbf{b}}$. Meanwhile, in the high field phase Aliouane {\it et al.} measured $M_b^{Tb}=3.92\,\mu_B$ of a total $M^{Tb}=7.24\,\mu_B$ at $T=8.5$~K and $H=5$~T\cite{AliouanePRL}. If we extrapolate from this, considering that his total moment is not far from saturation, to our conditions at $T=14$~K and $H=10$~T, we may assume a value of $M_b^{Tb}=4.15\pm0.35\,\mu_B$ at $10$~T, which would return a value of $A_C/H=-0.0061\pm0.0005$/T. The actual value from our fits in Fig.~\ref{fig:stokesH10} (see Table \ref{tab:ExpStrucFac}) is $A_C/H=-0.0060\pm0.0004$~/T, thus confirming that, as far as the $\Gamma_4$ part is concerned, the same magnetostrictive mechanism applies in the high field as well as in low field, i.e. in this phase we might also expect the presence of the Tb displacement along $\hat{\mathbf{c}}$ along with a similar oxygen displacement arising due to exchange striction\cite{WalkerScience}.

The second interesting observation, obtained from our fits to the Stokes measurements performed in both the incommensurate and commensurate magnetic phases, is that the relative populations of the clockwise and anticlockwise cycloidal domains are conserved on passing through the polarisation flop transition. Going from low to high field, the plane of the cycloid is rotated by $+90$ degrees about the $+\hat{\mathbf{b}}$ axis, for a magnetic field along $-\hat{\mathbf{b}}$. Then it turns back again when the field is returned to zero, with the same initial dominant domain restored with the same proportion. Such a behaviour was already seen in other experiments\cite{KimuraTMO,KimuraPRB}. This is somewhat surprising since the symmetry of the system does not favour one or the other domain of the new phase, in equilibrium with some domain of the initial phase. And if some dynamical effect would be considered at the transition, the domain opposite to the initial one would be preferred when returning to zero field. The answer to this conundrum may lie in the insufficiently precise alignment of the magnetic field with the $\hat{\mathbf{b}}$ axis. Indeed it has been shown\cite{Abe09} that a tilt of the magnetic field by $2$ degrees in the direction of $+\hat{\mathbf{a}}$ or $-\hat{\mathbf{a}}$ could significantly favour one or the other domain in the phase with ferroelectric polarisation parallel to $\hat{\mathbf{a}}$. Such an angle was within the uncertainty of our set up.

%
%

\section{CONCLUSIONS}
In conclusion, we have presented crystallographic and magnetic structural results for TbMnO$_3$, for a  magnetic field applied along $\hat{\mathbf{b}}$ and an electric field along $\hat{\mathbf{c}}$, using circularly polarised X-ray non-resonant scattering; bringing together an in-depth analysis of published results with previously unseen data. Starting from a discussion of circularly polarised X-ray non-resonant scattering for a generic spin cycloid, we demonstrated in detail how polarisation analysis reveals the $\mathbf{b}-\mathbf{c}$ cycloid in TbMnO$_3$ in zero magnetic field. This also highlights the importance of applying an electric field to produce a quasi-mono-domain state. Then for $0<H<H_C$ we explained how charge scattering interfering with the non-resonant magnetic scattering allows the determination of specific atomic displacements with femto-metre accuracy, revealing the connection between the electric polarisation and the magnetostrictive mechanisms. Finally we have presented new data in the polarisation flop phase, which indicates an interference between charge and magnetic scattering from a Mn $\mathbf{a}-\mathbf{b}$ cycloid, dominated by the charge. It was nevertheless possible to scale the charge structure factor with respect to the magnetic one, as in the low field phase, and to determine the cycloidal domain population.Our analysis shows that the Tb displacement along $\hat{\mathbf{c}}$ is similar in both multiferroic phases, and that the domain populations are preserved on passing through the magnetic field induced phase transition.

\begin{acknowledgements}
Thanks to everyone who helped with the experiments,
especially A. Fondacaro, J. Herrero-Martin, C. Mazzoli, G. Pepellin, V. Scagnoli, and T. Trenit. We also thank A. Malashevich for enlightening discussions, and T. Kimura for bringing valuable information to our attention.
\end{acknowledgements}

\section*{Appendix I}

 In this appendix details are given for making a qualitative comparison between the $H=0$~T data and the scattering expected for the model structure factor, similar to that employed in Section III's Eq.~\eqref{eq:H0kenz}, while adding in the Terbium $b$ component, with $S_b$=$({-i\hbar\omega}/{mc^2})s_b^{Tb}\cos(8\pi\Delta_a^{Tb})$, $L_b$=$({-i\hbar\omega}/{mc^2})l_b^{Tb}\cos(8\pi\Delta_a^{Tb})$

\begin{equation}\label{eq:channels0}
\begin{split}
F_{\sigma^\prime}=&M-i\epsilon\beta(\upsilon^\prime+i\alpha\upsilon^{\prime\prime})T_a+i\beta(\nu^\prime+i\alpha \nu^{\prime\prime})S_b,\\
F_{\pi^\prime}=&i\epsilon M-\beta(\upsilon^\prime+i\alpha\upsilon^{\prime\prime})T_a-\epsilon\beta(\nu^\prime+i\alpha \nu^{\prime\prime})(S_b+L_b)
\end{split}
\end{equation}

\noindent where and $\nu^\prime$+$i\alpha \nu^{\prime\prime}$=$e^{i\alpha\phi^{TM}_b}(e^{i\alpha\phi^{TO}_b}\!-\!1)$, such that:

\begin{equation}\label{eq:P2channels0}
\begin{split}
P_2\propto & |F_{\sigma^\prime}+F_{\pi^\prime}|^2-|F_{\sigma^\prime}-F_{\pi^\prime}|^2  =
\\
= &-8\beta \upsilon^\prime MT_a - 4\beta\epsilon \nu^\prime ML_b +  \\
  +  & 4\alpha (-\upsilon^{\prime\prime}\nu^\prime + \upsilon^\prime\nu^{\prime\prime})T_a(2S_b+L_b).
\end{split}
\end{equation}

The experiment shows that when the signs of $k$, $l$ and incident polarisation, that is $\alpha$, $\beta$ and $\epsilon$, are simultaneously reversed, $P_2$ is left qualitatively unchanged (see Fig.~\ref{fig:H0stokes}). This implies that in Eq.~\eqref{eq:P2channels0} the first and third terms cancel each other, the second one being non zero. Since the same is true after a sign change of $\beta$ alone, the first and third terms should both be negligible, that is $\upsilon^\prime\approx\upsilon^{\prime\prime}\approx 0$, $\nu^\prime\ne 0$. This in turn sets $\phi^{TO}_a$ to $\pi$, a value compatible with Kenzelmann's result\cite{kenzelmann}. $\phi^{TO}_b$ differs from $\phi^{TO}_a$ by either $0$ or $\pi$, but only $\phi^{TO}_b=\pi$ gives $\nu^\prime\ne 0$. With these phases our experiment is insensitive to the $a$ magnetic component of Tb and to the phase difference $\phi^{TM}_a$. To maximise $|\nu^\prime|$ requires that $\phi^{TM}_b=0,\pi$.

\section*{Appendix II}

We have used the $Pbnm$ setting for space group $\#62$, and according to the ITC notation for naming symmetry operations, they are as follows:
(1) 1, (2) 2($\frac{1}{2}$,0,0) $x,\frac{1}{4},0$, (3) 2(0,0,$\frac{1}{2}$) $0,0,z$, (4) 2(0,$\frac{1}{2}$,0) $\frac{1}{4},y,\frac{1}{4}$, (5) $\bar{1}$ $0,0,0$, (6) $b$ $\frac{1}{4},y,z$, (7) $m$ $x,y,\frac{1}{4}$ and (8) $n$($\frac{1}{2}$,0,$\frac{1}{2}$) $x,\frac{1}{4},z$.

The phase of the displacements is defined relative to the maximum in the $b$ component of the Mn magnetization. Consider a glide plane $n$ (operation 8 in the ITC Tables for $Pbnm$) at this maximum, it is a mirror, either even or odd, both for the magnetic field applied along $\hat{\mathbf{b}}$, and for the magnetic structure, and hence for the displacements. The field induced magnetisation $\Delta\mathbf{m}$ is invariant under the glide, but all components of the magnetic modulation, $\mathbf{m}(\mathbf{x})$ at $\mathbf{x}$, will be reversed:
\begin{equation}
n\Delta\mathbf{m}=\Delta\mathbf{m},\quad n\mathbf{x}=\mathbf{x}^\prime,\quad n\mathbf{m}(\mathbf{x})=-\mathbf{m}(\mathbf{x}^\prime),
\end{equation}
with the consequence for the displacement $\boldsymbol{\delta}(\mathbf{x})$, proportional to $\mathbf{m}\Delta\mathbf{m}$
\begin{equation}
n\boldsymbol{\delta}(\mathbf{x})=-\boldsymbol\delta(\mathbf{x}^\prime).\label{eqn:nglide}
\end{equation}
For the displacement modes, $n$ exchanges $1\!\!\leftrightarrow\!\!8$, $2\!\!\leftrightarrow\!\!7$, $3\!\!\leftrightarrow\!\!6$ and $4\!\!\leftrightarrow\!\!5$, and hence $\Delta_\alpha\leftrightarrow\varepsilon\Delta_\alpha$, $\Delta_\beta\leftrightarrow-\varepsilon\Delta_\beta$, $\Delta_\gamma\leftrightarrow-\varepsilon\Delta_\gamma$ and $\Delta_\delta\leftrightarrow\varepsilon\Delta_\delta$, where $\varepsilon$ is $+1$ for the $a$, $c$ components and $-1$ for $b$. If the sign agrees with that in Eqn~\eqref{eqn:nglide} then the displacement modulation is in phase with the magnetic one, and is labelled real. Whereas if the signs disagree, this implies that the displacement modulation is zero rather than a maximum at the maximum in m$_b^\mathrm{Mn}$, and is labelled as imaginary.

The $b$ glide operation exchanges positions: $1\!\!\leftrightarrow\!\!6$, $2\!\!\leftrightarrow\!\!5$, $3\!\!\leftrightarrow\!\!8$, and $4\!\!\leftrightarrow\!\!7$, giving $\Delta_\alpha\leftrightarrow\varepsilon\Delta_\alpha$, $\Delta_\beta\leftrightarrow\varepsilon\Delta_\beta$, $\Delta_\gamma\leftrightarrow-\varepsilon\Delta_\gamma$ and $\Delta_\delta\leftrightarrow-\varepsilon\Delta_\delta$, where $\varepsilon$ is $-1$ for the $a$ component and $+1$ for $b$ and $c$; whilst the action of screw axis $2y$ exchanges differently the mode components. By combining these two sets, and using the table for $G_k$, each component of each mode can be assigned to one specific irrep.

Finally, whether a mode is visible at ($4$, $\pm\tau$, $\pm1$) or not depends on the action of the $n$ glide. Since the structure factor is calculated for $F(h,0,l)$ where $h+l$ is odd, only the modes which change their sign under the action of the glide on the position, irrespective of the value of the factor $\varepsilon$, are visible, i.e. $\Delta_{\beta\pm}$ and $\Delta_{\gamma\pm}$. When the geometrical structure factors of the visible modes are calculated, with the phases found above, they happen to be purely real for modes in $\Gamma_4$ and purely imaginary for modes in $\Gamma_1$ and $\Gamma_2$, the latter considered in Section~\ref{s:H10}.

\bibliography{tmo_long}

\begin{thebibliography}{48}
\bibitem [{com()}]{comment}%
  \BibitemOpen
  \href@noop {} {}\bibinfo {note} {Note that this definition corresponds to
  $P_1$=$P_{\zeta}$, $P_2$=$P_{\xi}$ and $P_3$=$P_{\eta}$ in Ref.\cite{Blume},
  and that the electric field of the incident circular $\hat\epsilon_{c}^{+}$
  light polarization rotates clockwise for an observer looking towards the
  source.}\BibitemShut {Stop}%
\bibitem [{not({\natexlab{a}})}]{notedef}%
  \BibitemOpen
  \href@noop {} {} ({\natexlab{a}}),\ \bibinfo {note} {note this is the
  definition used in magnetic scattering, whilst the convention in
  crystallography is $\mathbf{K}=\mathbf{k}^\prime-\mathbf{k}$.}\BibitemShut
  {Stop}%
\bibitem [{not({\natexlab{b}})}]{noteCR}%
  \BibitemOpen
  \href@noop {} {} ({\natexlab{b}}),\ \bibinfo {note} {it should be noted that
  this low count rate is partially also a consequence of the use of the diamond
  phase plate, which at $E=6.16$~keV attenuates the incident photon flux by a
  factor of $\sim20$.}\BibitemShut {Stop}%
\end{thebibliography}

\end{document}